\documentclass[aps,prl,floatfix,showpacs,twocolumn,groupedaddress]{revtex4}
\usepackage[english]{babel}
\usepackage{graphicx}
\usepackage{dcolumn}
\usepackage{amsmath}

\newcommand{\eg}{{e.g.}}

\newcommand{\etal}{{\it et al.}}
\newcommand{\aBEDTTTFI}{$\alpha$-(BEDT-TTF)$_2$I$_3$}
\newcommand{\aETI}{$\alpha$-ET$_2$I$_3$}

\begin{document}

\title{Collective Excitations in the Charge-Ordered Phase of $\alpha$-(BEDT-TTF)$_2$I$_3$}

\author{T.\ Ivek}
\email{tivek@ifs.hr}
\homepage{http://real-science.ifs.hr/}
\author{B.\ Korin-Hamzi\'{c}}
\author{O.\ Milat}
\author{S.\ Tomi\'{c}}
\affiliation{Institut za fiziku, P.O.Box 304, HR-10001 Zagreb, Croatia}
\author{C.\ Clauss}
\author{N.\ Drichko}
\author{D.\ Schweitzer}
\author{M.\ Dressel}
\affiliation{Physikalisches Institut, Universit\"{a}t Stuttgart, D-70550, Stuttgart, Germany}

\date{\today}

\begin{abstract}
The charge response of charge-ordered state in the organic 
conductor $\alpha$-(BEDT-TTF)$_2$I$_3$ is characterized by dc resistivity, 
dielectric and optical spectroscopy in different crystallographic directions 
within the two-dimensional conduction layer. Two dielectric modes are detected. The 
large mode is related to the phason-like excitation of the 
$2k_{\mathrm{F}}$ bond-charge density wave which forms in the $ab$ plane. 
The small dielectric mode is associated with the motion of domain-wall pairs 
along the $a$- and $b$-axes between two types of domains which are created due
to inversion symmetry breaking. 
\end{abstract}


%
%
%
%
%
%
%

\maketitle

Electron-electron and electron-phonon interactions, in particular in 
the systems with reduced dimensionality, are known to be the driving force for 
the formation of new ordered states of matter \cite{Fulde93,MC04}. Among the 
most intriguing phenomena found in these systems are broken symmetry phases like 
charge- and spin-density waves (CDW, SDW), charge order (CO), antiferromagnetic 
and spin-Peierls phases; all of them show a large variety of nonlinear 
properties and complex dynamics, including collective excitations 
\cite{Gruener88,Littlewood87,Cava85,VuleticPR}. This wealth of charge-ordered 
phenomena in one- (1D) and two-dimensional (2D) strongly correlated systems is 
brought in by the variety in lattice structure, represented by anisotropic 
networks, both in the electron hopping $t$ and in the inter-site Coulomb 
interaction $V$. While CO is an effect observed both in inorganic and 
organic materials with strong electronic correlations, the organics present a 
more robust and ``clean'' state, where in some cases CO does not 
compete with other ground states, as opposed to high-temperature 
superconducting materials, for instance \cite{Tranquada08}.

Theoretical considerations indicate that the charge disproportionation is driven 
by Coulomb repulsion. In particular, for systems with a quarter-filled 
conduction band even large values of on-site repulsion $U$ are not sufficient to 
transform the ground state from metallic to insulating; rather the inter-site 
Coulomb interaction $V$ is responsible to stabilize a Wigner-crystal type phase
(Seo \etal{}\ in Ref.\ \cite{OC06}). Indeed, there are several well-established
charge-ordered organic 1D and 2D systems \cite{OC06}:
charge modulation driven by long-range Coulomb 
repulsion was first evidenced in the 1D organic conductors (DI-DCNQI)$_2$Ag by 
Kanoda in Ref.\ \cite{OC06} and (TMTTF)$_2X$ \cite{Monceau01} and also in 2D 
conductors based on the BEDT-TTF [bis(ethylenedithio)tetrathiafulvalene] 
molecule: $\theta$-(BEDT-TTF)$_2$RbZn(SCN)$_4$ and \aBEDTTTFI{} by Takahashi 
\etal{}\ \cite{OC06}.

The organic conductor \aBEDTTTFI{} (\aETI{}) is the most prominent example of 
charge order in 2D organic conductors. The crystals are formed by alternating 
anion and donor layers in the $ab$ plane; the unit cell is triclinic and contains four BEDT-TTF 
molecules. The BEDT-TTF layer consists of two types of stacks: Stack I is 
weakly dimerized and composed of crystallographically equivalent molecules A and 
A$^{\prime}$, while the stack II is a uniform chain composed of B and C 
molecules \cite{Bender84}. At high temperatures the system is a semimetal with 
small electron and hole pockets at the Fermi surface \cite{Mori84}.
At the metal-to-insulator transition $T_{\mathrm{CO}} = 136$~K  the conductivity
drops by several orders of magnitude and a temperature-dependent gap opens in
the charge and spin sector which indicates the insulating and diamagnetic nature
of the ground state \cite{Bender84}. Nuclear magnetic resonance
(NMR) \cite{Takano01} and synchrotron X-ray diffraction measurements \cite{Kakiuchi07}
demonstrate that the charge order, whose fluctuations are already observed at
high temperatures \cite{Moroto04}, develops at extended length scales below
$T_{\mathrm{CO}}$. The estimated charge values of the molecules are
$\mathrm{A}=0.82(9)$, $\mathrm{A}^{\prime}=0.29(9)$, $\mathrm{B}=0.73(9)$ and
$\mathrm{C}=0.26(9)$. The exact site assignment is not settled yet since these
values slightly differ from those found in the NMR, vibrational infrared and
Raman spectroscopy \cite{Takano01,Moldenhauer93,Wojciechowski03}. Observed
structural changes comprise two effects. First is a symmetry reduction at the
phase transition with the space group $P\overline{1}$ changing into $P1$ in the
CO state, which implies four non-equivalent BEDT-TTF molecules in the unit cell.
Second is the modulation of overlap integrals due to changes in dihedral angles
at low temperatures. In other words, the charge order comprises ``horizontal''
charge stripes along the $b$ crystallographic axis of charge-poor (CP) sites,
the A$^{\prime}$ and C molecules, and charge-rich (CR) sites, the A and B
molecules, together with a bond modulation. The latter infers that the
nature of CO in this system should not be regarded as fully localized;
instead, a CDW picture would be more suitable.

\begin{figure*}
\includegraphics[clip,width=1.0\linewidth]{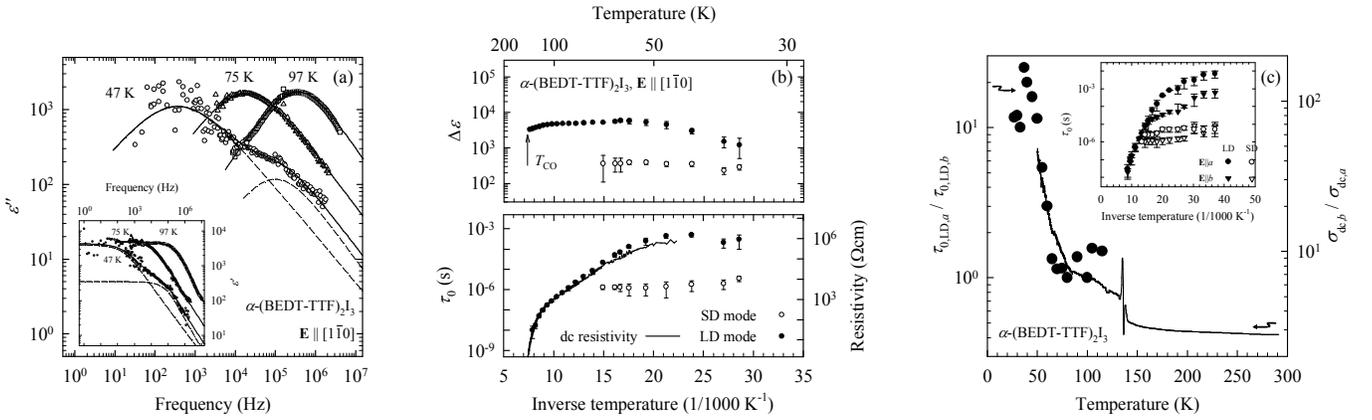}
\caption{\label{fig1}(a) Frequency dependence of the real ($\varepsilon^{\prime}$) 
(inset) and imaginary ($\varepsilon^{\prime\prime}$) (main panel) part of the 
dielectric function in \aETI{} at representative temperatures for 
$\mathbf{E}\parallel [1 \bar{1} 0]$. Below 75~K two relaxation modes are 
observed -- full lines for 47~K show a fit to a sum of two generalized Debye 
functions; dashed lines represent contributions of the two modes. Above 75~K 
only one mode is detected, and the full lines represent fits to a single 
generalized Debye function.
(b) Dielectric strength (upper panel) and mean relaxation time
with dc resistivity (points and line, respectively, lower panel) in \aETI{}
as a function of inverse temperature, for $\mathbf{E}\parallel [1 \bar{1} 0]$.
(c) Anisotropy of large dielectric mode mean relaxation times 
in \aETI{} (points, left axis). The temperature behavior closely follows 
the dc conductivity anisotropy (solid line, right axis). In the inset the mean 
relaxation time is plotted as a function of inverse temperature; full and empty 
symbols mark the large and small dielectric mode, respectively, for $\mathbf{E}$ 
along the $a$- (circles) and $b$-axis (triangles).}
\end{figure*}

The considerable knowledge of the CO pattern in \aETI{} makes it a 
suitable candidate in the search for collective excitations in this CO phase.
Early results indicated the existence of a broad relaxation in radio-frequency
range with a large dielectric constant of the order of $10^5$ as well as
sample-dependent nonlinearities \cite{Dressel95}.
In an attempt to characterize the collective excitations of the charge order, we
have undertaken dc and ac conductivity-anisotropy measurements on carefully
oriented single crystals of \aETI{}. We discovered a complex and
anisotropic dispersion in the charge-ordered state of a 2D organic crystal.
First, similar to the Peierls CDW  state, we observe long-wavelength charge
excitations with an anisotropic phason-like dispersion, which we detect as broad
screened relaxation modes along both the $a$- and $b$-axis of the 2D BEDT-TTF
plane. Second, we detect short  wavelength charge excitations in the form of
domain wall pairs, created by breaking the inversion symmetry, which are less
mobile and induce a much weaker  polarization, again along both crystallographic
axes. Also, we observe both  types of excitations along diagonal direction of
BEDT-TTF plane. Our results are well understood within the theoretical model
by Clay \etal{}\ \cite{Clay02} which shows that the CO phase with horizontal
stripes of localized charges is best described in terms of a bond-charge density
wave.

The dc resistivity of \aETI{} was measured between room temperature and 40~K. In the
frequency range 0.01~Hz--10~MHz the complex dielectric function was 
obtained from the complex conductance measured by two setups. At high 
frequencies (40~Hz--10~MHz) an Agilent 4294A precision impedance analyser was 
used. At low frequencies (0.01~Hz--3~kHz) a setup for measuring high-impedance 
samples was used based on lock-in technique. At frequencies 6--10000~cm$^{-1}$ 
the complex dielectric function was obtained by a Kramers-Kronig analysis of the 
infrared reflectivity. All experiments were done on flat, planar high-quality 
single crystals along three directions within the plane: in the $a$ and $b$ 
crystallographic axes as well as in $[1 \bar{1} 0]$ (diagonal) direction 
\cite{note00}. An influence of extrinsic effects in the dielectric 
measurements, especially those due to contact resistance and surface layer 
capacitance, was ruled out with scrutiny \cite{Ivek08}.

Fig.\ \ref{fig1}(a) shows representative spectra at three selected temperatures for 
$\mathbf{E}\parallel [1 \bar{1} 0] $ taken below $T_{\mathrm{CO}} = 136.2$~K in 
the charge-ordered state. Most notably, between 35~K and up to 75~K two 
dielectric relaxation modes are discerned. The complex dielectric spectra 
$\varepsilon(\omega)$ can be described by the sum of two generalized Debye 
functions $\varepsilon(\omega)-\varepsilon_\infty
 = \Delta\varepsilon_{\mathrm{LD}} / [ 1 + \left(i \omega \tau_{0,\mathrm{LD}} \right)^{ 1-\alpha_{\mathrm{LD}} } ]
 + \Delta\varepsilon_{\mathrm{SD}} / [ 1 + \left(i \omega \tau_{0,\mathrm{SD}} \right)^{ 1-\alpha_{\mathrm{SD}} } ]$
where $\varepsilon_\infty$ is the high-frequency dielectric constant, 
$\Delta\varepsilon$ is the dielectric strength, $\tau_0$ the mean relaxation 
time and $1-\alpha$ the symmetric broadening of the relaxation time distribution 
function of the large (LD) and small (SD) dielectric mode. The broadening 
parameter $1-\alpha$ of both modes is typically $0.70 \pm 0.05$. In Fig.\ \ref{fig1}(b)
the dielectric strengths and mean relaxation times are plotted as a function of 
inverse temperature. The dielectric strength of both modes does not change 
significantly with temperature ($\Delta\varepsilon_{\mathrm{LD}} \approx 5000$, 
$\Delta\varepsilon_{\mathrm{SD}} \approx 400$). At approximately 75~K the large 
dielectric mode overlaps the small mode. It is not clear whether the small 
dielectric mode disappears at this temperature or is merely obscured by the 
large dielectric mode due to its relative size. However, above 100~K, when the 
large dielectric mode shifts to high enough frequencies, no indication can be 
found of a smaller mode centered in the range $10^5$--$10^6$~Hz. Accordingly, 
above 75~K fits to only one generalized Debye function are performed which we
identify with the continuation of the large dielectric mode. All three of the
large-mode parameters can be extracted in full detail until it exits our
frequency window at about 130~K. At temperatures up to 135~K (just below
$T_{\mathrm{CO}}$) we can determine only the dielectric relaxation 
strength by measuring the capacitance at 1~MHz.

The most intriguing result is that the dielectric strength and temperature
behavior of the mean relaxation times differ greatly between the two dielectric
modes. The large one follows a thermally activated behavior similar
to the dc resistivity, whereas the small mode is almost
temperature-independent. For both orientation $\mathbf{E}\parallel a$ and 
$\mathbf{E}\parallel b$ the results are comparable to the findings along the 
$\mathbf{E}\parallel [1 \bar{1} 0]$. There is no pronounced anisotropy 
or temperature dependence in the dielectric strength: the $\Delta\varepsilon$ 
values of both the large and small dielectric modes (not shown) correspond to 
those measured with $\mathbf{E}\parallel [1 \bar{1} 0]$. However, an anisotropy 
in $\tau_{0,\mathrm{LD}}$ is clearly visible and its evolution closely follows 
the dc conductivity anisotropy [Fig.\ \ref{fig1}(c)]. A similar conductivity 
anisotropy has been observed in the CO phase of (TMTTF)$_2$AsF$_6$ 
\cite{KorinHamzic06}. In our samples, despite the temperature-dependent 
activation, the anisotropic transport gap in the CO phase for 
$\mathbf{E}\parallel a$ and  $\mathbf{E}\parallel b$ can be estimated to about 
$2\Delta = 80$~meV and 40~meV. Conversely, our optical measurements for 
$\mathbf{E}\parallel a$ and $\mathbf{E}\parallel b$ reveal the isotropic gap of 
about 75~meV \cite{Clauss09}. This is not surprising since systems with complex 
band structure can exhibit distinct optical and transport 
gaps: optical measurements probe direct transitions between the valence and 
conduction band, while dc transport is governed by transitions with the smallest 
energy difference between the two bands.

\begin{figure}
\includegraphics[clip,width=0.7\linewidth]{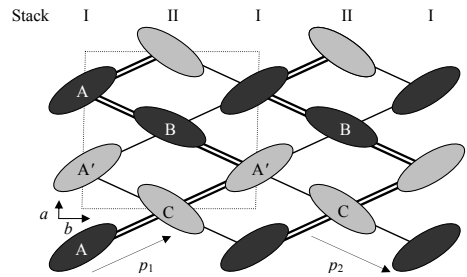}
\caption{\label{fig2}Schematic representation of a $2k_{\mathrm{F}}$ bond-CDW in 
\aETI{}. Thin dotted line denotes a unit cell. Single and double lines represent
modulated overlap integrals along two $p$-directions (see text).}
\end{figure}

The observed ac conductivity data demonstrate a complex and anisotropic 
dispersion in the CO state. First, similar to CDW state, 
we observe broad screened relaxation (large dielectric) mode along the $a$- and 
$b$-axis and for $\mathbf{E}\parallel [1 \bar{1} 0] $. This fact indicates that the
observed dielectric relaxation is not associated with stripe orientation and its
motion along one of the two in-plane crystallographic axes. Rather, this mode can
be interpreted as a signature of long-wavelength CDW excitations possessing an 
anisotropic phason-like dispersion. A 2D dispersion with similar dielectric
strength amplitude and relaxation time behavior was previously found 
in the commensurate CDW phase developed in ladder layers of Sr$_{14}$Cu$_{24}$O$_{41}$ 
\cite{VuleticPRL,Vuletic05,Abbamonte04}. In \aETI{} Kakiuchi \etal{}\ suggested
a $2k_{\mathrm{F}}$ CDW formation along the zig-zag path CABA$^\prime$C of large
overlap integrals detected in their X-ray diffraction measurements \cite{Kakiuchi07}.
A theoretical model for the related quarter-filled
$\theta$-ET$_2$X systems may provide additional insight \cite{Clay02}.
Clay \etal{} showed that the CO phase with horizontal stripe phase is
characterized by a 1100 modulation of site charges along the two independent
$p$-directions parallel to the larger overlap integral, 
and with a 1010 modulation along the $b$ axis perpendicular to the stripes. 
In addition, this CO is
accompanied by a modulation of the overlap integrals along the same $p$-direction.
In other words, such a CO phase corresponds to the $2k_{\mathrm{F}}$ modulation
of bonds and site charges, \eg{}, a combined bond-CDW along the two BEDT-TTF plane $p$-directions. An analogous, albeit more complex,
$2k_{\mathrm{F}}$ modulation of overlap integrals indeed develops along the
$p$-directions of \aETI{}, ACA$^\prime$BA
and ABA$^\prime$CA (see Fig.\ \ref{fig2}) \cite{Kakiuchi07}. It is plausible to look for the origin
of phason-like dielectric relaxation in such a $2k_{\mathrm{F}}$ bond-CDW. In this
case the energy scale of barrier heights is close to the single-particle
activation energy indicating that screening by single carriers responsible for 
the dc transport is effective for this relaxation. The fact that the 
temperature behavior of $\tau_{0,\mathrm{LD}}$ anisotropy closely follows the dc 
conductivity anisotropy has important implications: while the CDW motion is 
responsible for the dielectric response, the single electron/hole motion along
the two $p$-directions, possibly zig-zagging between them, is responsible for
the observed dc charge transport.

\begin{figure}
\includegraphics[clip,width=0.9\linewidth]{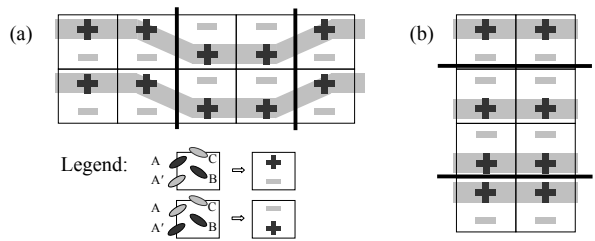}
\caption{\label{fig3}Two different types of domain wall pairs in the
charge-ordered phase of \aETI{}. (A,B)- and (A$^{\prime}$,B)-rich unit
cells are symbolically represented as $+-$ or $-+$ cells which form CO stripes.
For simplicity we omit the B and C molecules. Gray thick lines stand for
charge-rich stripes. Thin black lines denote a domain wall pair.}
\end{figure}

Second, we address the small dielectric mode. Here we note the twinned nature of
the CO  phase due to breaking the inversion symmetry and formation of a 1010
modulation of site charges along the $a$-axis, with one domain being (A,B)-rich
and the other (A$^{\prime}$,B)-rich \cite{Kakiuchi07}. Indeed, the 
ferroelectric aspect of the CO phase was also theoretically suggested \cite{OC06} and experimentally probed \cite{Yamamoto08}.
Our data can be naturally attributed to the motion of charged kink-type 
defects -- solitons or domain walls in the CO texture.
Charge neutrality constraint of the CO in \aETI{} (a change of stripes 
equivalent to strictly replacing unit cells of one twin type with another) 
suggests two types of solitons and/or domain walls. The first one is the domain
 wall in pairs (a soliton-antisoliton pair) between CR and CP stripes along the 
$b$-axis, which we get if we impose the constraint along the $b$-axis
[Fig.\ \ref{fig3}(a)]. The second type of domain-wall pair is given by applying
the constraint along the $a$-axis in such a way that the domain walls' interior contains both
charge signs [Fig.\ \ref{fig3}(b)]. The motion of such entities induces a 
displacement current and can therefore be considered as the microscopic origin 
of polarization in the CO state. Their relaxation, being nearly
temperature-independent, cannot be dominated by resistive dissipation, rather it
is governed by low-energy barriers, similarly as observed in ferromagnetic domain
phase \cite{Pinteric99}. 
Finally, the observed dielectric strength $\Delta\varepsilon$ of the small mode of about 400
confirms the domain-wall assignment since their dynamics, in contrast to
phason-like one, is commonly found to be characterized by much smaller
dielectric constants (of the order 1000 and less) \cite{Tokura89,Tomic01}.

In conclusion, we demonstrate the development of in-plane anisotropic
conductivity accompanied by the appearance of two low-frequency dielectric
relaxation modes in the charge-ordered phase with a horizontal stripe structure of
\aBEDTTTFI{}. The large dielectric mode features an anisotropic phason-like
behavior which we associate with the $2k_{\mathrm{F}}$ bond-charge density wave
formed within the two $p$-directions in the BEDT-TTF plane. On the other hand, 
we ascribe the small dielectric mode to the motion of domain-wall pairs
at interfaces between two different types of charge order domains created due to
inversion symmetry breaking. This unusual appearance of a phason-like relaxation 
alongside a soliton-like mode underlines the complexity of collective
excitations in such a charge order system. Our results strongly suggest that a
localized Wigner-like picture in terms of stripes of localized site charges is
not appropriate for charge order in \aBEDTTTFI{}. Instead, our results are
consistent with a description in terms of a 2D bond-charge density wave. Our
findings demonstrate that the issue of collective excitations in broken symmetry
phases becomes ever more subtle and call for further work in order to clarify
whether such a dispersion is common to 2D charge order phases with horizontal
stripes in diverse strongly correlated systems.

\begin{acknowledgments}
We acknowledge S.\ Mazumdar for valuable and illuminating discussions. We thank
G.\ Untereiner for the sample preparation and T.\ Vuleti\'{c} for his
help in data analysis. This work was supported by the Croatian Ministry of
Science, Education and Sports under grants 035-0000000-2836 and
035-0352843-2844 and by the Deutsche Forschungsgemeinschaft under grant
DR 228/29-1.
\end{acknowledgments}


\begin{thebibliography}{99}
\bibitem{note00}
Single crystals were oriented in the mid-infrared set-up before,
and in the X-ray set-up after experiments.
\end{thebibliography}
\end{document}